%% file: paper.tex
\documentclass[preprint]{sigplanconf}

\usepackage{amsmath}
\usepackage{url}
\usepackage{graphicx}

\usepackage{tikz}
\usepackage{tkz-graph}
\usetikzlibrary{arrows}
\usetikzlibrary{fit}					% fitting shapes to coordinates
\usetikzlibrary{backgrounds}	% drawing the background after the foreground
\usetikzlibrary{arrows}

\usepackage{color}
\definecolor{Red}{rgb}{0.9,0.0,0.0}  % fixme
\definecolor{Green}{rgb}{0.0,0.4,0.0}
\definecolor{Blue}{rgb}{0.0,0.0,0.9}
\definecolor{DarkBlue}{rgb}{0.0,0.0,0.75}
\definecolor{Midnight}{rgb}{0.0,0.0,0.5}
\definecolor{Purple}{rgb}{0.5,0.0,0.4}
\definecolor{Black}{rgb}{0.0,0.0,0.0}
\definecolor{Yellow}{rgb}{1.0,1.0, 0.25}
\definecolor{Cyan}{rgb}{0.25,1.0, 1.0}

\newcommand{\cdColor}{Black}
\newcommand{\kwColor}{DarkBlue}
\newcommand{\comColor}{Red}

\usepackage{listings}
\usepackage{textcomp}
\input{lstlang-rust}
\usepackage{algorithm}
\usepackage{algpseudocode}

\input{common-defs}

\usepackage{code}

\title{Experience Report: Developing the Servo Web Browser Engine using Rust}

\authorinfo{%
  Brian Anderson \and Lars Bergstrom \\
  David Herman \and Josh Matthews \\
  Keegan McAllister \and Jack Moffitt \\
  Simon Sapin
}{%
  Mozilla Research}{%
  \{banderson,larsberg,dherman,jdm,\\kmcallister,jack,ssapin\}@mozilla.com}
\authorinfo{%
  Manish Goregaokar
}{Indian Institute of Technology Bombay}{%
  manishg@iitb.ac.in}
\date{\today}

\begin{document}

\maketitle
\thispagestyle{empty}

\sloppy

\input{abstract}
\input{intro}
\input{browsers}
\input{rust}
\input{servo}
\input{open}
\input{rel}

\acks
The Servo project was started at Mozilla Research, but has benefitted from
contributions from Samsung, Igalia, and hundreds of volunteer community members.
These contributions amount to more than half of the commits to the project,
and we are grateful for both our partners' and volunteers' efforts to make
this project a success.

\bibliographystyle{alpha}
\bibliography{strings-short,manticore,browser}

\end{document}

%% file: lstlang-rust.tex
\lstdefinelanguage{Rust}[]{C}{%
  morekeywords={abstract,alignof,as,become,box,break,const,continue,crate,do,enum,extern,false,
  final,fn,for,if,impl,in,let,loop,macro,match,mod,move,mut,offsetof,override,priv,pub,pure,ref,return,
  sizeof,static,self,struct,super,true,trait,type,typeof,unsafe,unsized,use,virtual,where,while,yield},
  otherkeywords={::,->,\#},
}[keywords,comments,strings]%

%% file: common-defs.tex
\usepackage{times}
\SetMathAlphabet{\mathtt}{normal}{OT1}{pcr}{m}{n}
\SetMathAlphabet{\mathtt}{bold}{OT1}{pcr}{bx}{n}
\usepackage{amsmath}
\usepackage{amssymb} % for \pitchfork

\newcommand{\CUT}[1]{}

\newcommand{\secref}[1]{Section~\ref{#1}}
\newcommand{\tblref}[1]{Table~\ref{#1}}
\newcommand{\figref}[1]{Figure~\ref{#1}}

\newcommand{\Cplusplus}{\mbox{C\hspace{-.05em}\raisebox{.4ex}{\tiny\bf ++}}}

\newcommand{\ocaml}{\textsc{OCaml}}

\providecommand{\bftt}[1]{{\ttfamily\bfseries{}#1}}

\providecommand{\kw}[1]{\bftt{#1}}
\providecommand{\nt}[1]{{\rmfamily\itshape{#1}}}

\newcount\timeHH
\newcount\timeMM
\timeHH=\time
\divide\timeHH by 60
\timeMM=\time
\multiply\count255 by -60 \advance\timeMM by \count255
\newcommand{\timestamp}{%
  \today{} ---
  \ifnum\timeHH<10 0\fi\number\timeHH\,:\,\ifnum\timeMM<10 0\fi\number\timeMM}

%% file: abstract.tex
%!TEX root = paper.tex
%
\begin{abstract}
All modern web browsers --- Internet Explorer, Firefox, Chrome, Opera, and Safari ---
have a core rendering engine written in \Cplusplus{}.
This language choice was made because it affords the systems programmer complete control
of the underlying hardware features and memory in use, and it provides a transparent
compilation model.

Servo is a project started at Mozilla Research to build a new web browser engine that
preserves the capabilities of these other browser engines but also both takes 
advantage of the recent trends in parallel hardware and is more memory-safe.
We use a new language, Rust, that provides us a similar level of control of the underlying
system to \Cplusplus{} but which builds on many concepts familiar to the functional programming
community, forming a novelty --- a useful, safe systems programming language.

In this paper, we show how a language with an affine type system, regions, and many syntactic
features familiar to functional language programmers can be successfully used to build
state-of-the-art systems software.
We also outline several pitfalls encountered along the way and describe some potential
areas for future research.
\end{abstract}

%% file: intro.tex
%!TEX root = paper.tex
\section{Introduction}
\label{sec:intro}
The heart of a modern web browser is the browser engine, which is the code responsible
for loading, processing, evaluating, and rendering web content.
There are three major browser engine families:
\begin{enumerate}
\item Trident/Spartan, the engine in Internet Explorer~\cite{IE}
\item Webkit\cite{WEBKIT}/Blink, the engine in Safari~\cite{SAFARI}, Chrome~\cite{CHROME}, and Opera~\cite{OPERA}
\item Gecko, the engine in Firefox~\cite{FIREFOX}
\end{enumerate}
All of these engines have at their core many millions of lines of \Cplusplus{} code.
While the use of \Cplusplus{} has enabled all of these browsers to achieve excellent sequential
performance on a single web page, on mobile devices with lower processor speed but many
more processors, these browsers do not provide the same level of interactivity that they
do on desktop processors~\cite{parallelizing-web-pages,ZOOMM}.
Further, in an informal inspection of the critical security bugs in Gecko, we determined that
roughly 50\% of the bugs are use after free, out of range access, or related to integer
overflow.
The other 50\% are split between errors in tracing values from the JavaScript heap in the
\Cplusplus{} code and errors related to dynamically compiled code.

Servo~\cite{SERVO} is a new web browser engine designed to address the major environment and 
architectural changes over the last decade.
The goal of the Servo project is to produce a browser that enables new applications to be authored
against the web platform that run with more safety, better performance, and better power usage
than in current browsers.
To address memory-related safety issues, we are using a new systems programming language,
Rust~\cite{RUST}.
For parallelism and power, we scale across a wide variety of hardware by building either data-
or task-parallelism, as appropriate, into each part of the web platform.
Additionally, we are improving concurrency by reducing the simultaneous access to data
structures and using a message-passing architecture between components such as the
JavaScript engine and the rendering engine that paints graphics to the screen.
Servo is currently over 400k lines of Rust code and implements enough of the web platform to render and
process many pages, though it is still a far cry from the over 7 million lines of code in
the Mozilla Firefox browser and its associated libraries.
However, we believe that we have implemented enough of the web platform to provide an
early report on the successes, failures, and open problems remaining in Servo, from the
point of view of programming languages and runtime research.
In this experience report, we discuss the design and architecture of a modern web 
browser engine, show how modern programming language techniques --- many of which
originated in the functional programming community --- address these design 
constraints, and also touch on ongoing challenges and areas of research where we
would welcome additional community input.

%%% Local Variables: 
%%% mode: latex
%%% TeX-master: "paper"
%%% End: 

%% file: browsers.tex
%!TEX root = paper.tex

\section{Browsers}
\label{sec:browsers}

\begin{figure*}[ht]
  \begin{center}
    \includegraphics[scale=0.55]{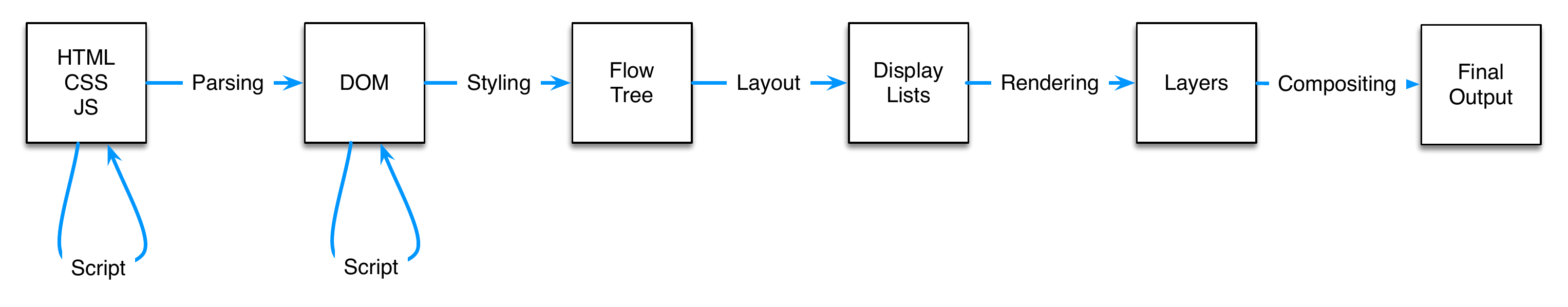}
  \end{center}%
  \caption{Processing stages and intermediate representations in a browser engine.}
  \label{fig:browser}
\end{figure*}

Modern web browsers do not just load static pages, but can also handle pages that have similar
complexity to native applications.
From application suites such as the Google Apps\footnote{\url{https://apps.google.com}} to games
based on the Unreal Engine,\footnote{\url{https://www.unrealengine.com/}}
modern browsers have the ability to handle much more than simple static pages.
\figref{fig:browser} shows the steps in processing a site.
While the naming is specific to the Servo browser, similar steps are used in all modern browsers.\footnote{\url{http://www.html5rocks.com/en/tutorials/internals/howbrowserswork/}}

\subsection{Parsing HTML and CSS}

A URL identifies a resource to load.
This resource usually consists of HTML, which is then parsed and typically turned into a Document Object
Model (DOM) tree.
From a programming languages standpoint, there are several interesting aspects of the parser design
for HTML.
First, though the specification allows the browser to abort on a parse error,\footnote{\url{https://html.spec.whatwg.org/multipage/#parsing}}
in practice browsers follow the recovery algorithms described in that specification precisely so that
even ill-formed HTML will be handled in an interoperable way across all browsers.
Second, due to the presence of the \lstinline[language=HTML]{<script>} tag, the token stream can be modified
during operation.
For example, the below example that injects an open tag for the header and comment blocks works in all modern browsers.
\begin{lstlisting}[language=HTML]
<html>
  <script>
  document.write("<h");
  </script>1>
  This is a h1 title

  <script>
  document.write("<!-");
  </script>-
  This is commented
  -->
</html>
\end{lstlisting}
This requires parsing to pause until JavaScript code has run to completion.
But, since resource loading is such a large factor in the latency of loading many webpages (particularly on mobile),
all modern parsers also perform speculative token stream scanning and prefetch of resources likely to be required~\cite{browsers-slow-smartphones}.

\subsection{Styling}

After constructing the DOM, the browser uses the styling information in linked
CSS files and in the HTML to compute a styled tree of flows and fragments,
called the \emph{flow tree} in Servo.
This process can create many more flows than previously existed in the DOM ---
for example, when a list item is styled to have an associated counter glyph.

\subsection{Layout}

The flow tree is then processed to produce a set of \emph{display list} items.
These list items are the actual graphical elements, text runs, etc. in their
final on-screen positions.
The order in which these elements are displayed is well-defined by the
standard.\footnote{\url{http://www.w3.org/TR/CSS21/zindex.html#painting-order}}

\subsection{Rendering}

Once all of the elements to appear on screen have been computed, these
elements are rendered, or painted, into memory buffers or directly to graphics
surfaces.

\subsection{Compositing}

The set of memory buffers or graphical surfaces, called \emph{layers}, are then
transformed and composited together to form a final image for
presentation. Layerization is used to optimize interactive transformations
like scrolling and certain animations.

\subsection{Scripting}

Whether through timers, \lstinline[language=HTML]{<script>} blocks in the
HTML, user interactions, or other event handlers, JavaScript code may execute
at any point during parsing, layout, and painting or afterwards during
display.  These scripts can modify the DOM tree, which may require rerunning
the layout and painting passes in order to update the output.  Most modern
browsers use some form of dirty bit marking to attempt to minimize the
recalculations during this process.

%%% Local Variables: 
%%% mode: latex
%%% TeX-master: "paper"
%%% End: 

%% file: rust.tex
%!TEX root = paper.tex

\section{Rust}
\label{sec:rust}

Rust is a statically typed systems programming language most heavily inspired by the C
and ML families of languages~\cite{RUST}.
Like the C family of languages, it provides the developer fine control over memory layout
and predictable performance.
Unlike C programs, Rust programs are \emph{memory safe} by default,
only allowing unsafe operations in specially-delineated blocks.

Rust features an \emph{ownership-based} type system inspired by the region systems work in the
MLKit project~\cite{mlkit} and especially as implemented in the Cyclone language~\cite{cyclone}.
Unlike the related ownership system in Singularity OS~\cite{singularity}, Rust allows programs to
not only transfer ownership but also to temporarily \emph{borrow} owned values, significantly
reducing the number of region and ownership annotations required in large programs.
The ownership model encourages immutability by default while allowing for controlled
mutation of owned or uniquely-borrowed values.

Complete documentation and a language reference for Rust are available at: \url{http://doc.rust-lang.org/}.

\subsection{Ownership and concurrency}
Because the Rust type system provides very strong guarantees about memory aliasing,
Rust code is memory safe even in concurrent and multithreaded environments,
but beyond that Rust also ensures \emph{data-race freedom}.

In concurrent programs, the data operated on by distinct threads is also itself distinct:
under Rust's ownership model, data cannot be owned by two threads at the same time.
For example, the code in \figref{fig:bad-concurrency} generates a static error from the compiler because
after the first thread is spawned, the ownership of \lstinline{data} has been transferred into the
closure associated with that thread and is no longer available in the original thread.
%http://is.gd/tkcD9R
\begin{figure}
\begin{lstlisting}
fn main() {
  // An owned pointer to a heap-allocated
  // integer
  let mut data = Box::new(0);

  Thread::spawn(move || {
    *data = *data + 1;
  });
  // error: accessing moved value
  print!("{}", data);
}
\end{lstlisting}
  \caption{Code that will not compile because it attempts to access mutable state from two threads.}
  \label{fig:bad-concurrency}
\end{figure}

On the other hand, the immutable value in \figref{fig:shared-concurrency} can be borrowed and shared between multiple threads as long as those threads don't outlive the scope of the data, and even mutable values can be shared as long
as they are owned by a type that preserves the invariant that mutable memory is unaliased, as with the
mutex in \figref{fig:shared-mutable-concurrency}.

\begin{figure}
\begin{lstlisting}
fn main() {
  // An immutable borrowed pointer to a
  // stack-allocated integer
  let data = &1;

  Thread::scoped(|| {
    println!("{}", data);
  });
  print!("{}", data);
}
\end{lstlisting}
  \caption{Safely reading immutable state from two threads.}
  \label{fig:shared-concurrency}
\end{figure}

\begin{figure}
\begin{lstlisting}
fn main() {
  // A heap allocated integer protected by an
  // atomically-reference-counted mutex
  let data = Arc::new(Mutex::new(0));
  let data2 = data.clone();

  Thread::scoped(move || {
    *data2.lock().unwrap() = 1;
  });

  print!("{}", *data.lock().unwrap());
}
\end{lstlisting}
  \caption{Safely mutating state from two threads.}
  \label{fig:shared-mutable-concurrency}
\end{figure}

With relatively few simple rules, ownership in Rust enables foolproof task parallelism,
but also data parallelism, by partitioning vectors and lending mutable references into properly scoped threads.
Rust's concurrency abstractions are entirely implemented in libraries, and though
many advanced concurrent patterns such as work-stealing~~\cite{blumeofe:multiprogrammed-work-stealing}
cannot be implemented in safe Rust, they can usually be encapsulated in a memory-safe interface.

%%% Local Variables: 
%%% mode: latex
%%% TeX-master: "paper"
%%% End: 

%% file: servo.tex
\section{Servo}
\label{sec:servo}
A crucial test of Servo is performance --- Servo must be at least as fast
as other browsers at similar tasks to succeed, even if it provides additional memory safety.
\tblref{servo-perf} shows a preliminary comparison of the performance of the layout stage (described
in \secref{sec:browsers}) of rendering several web sites in Mozilla Firefox's Gecko engine compared
to Servo, taken on a modern MacBook Pro.
\begin{table}
  \begin{center}
    \begin{tabular}{r || r | r | r}
      Site & Gecko & Servo 1 thread & Servo 4 threads \\
      \hline
      Reddit & 250 & 100 & 55  \\
      CNN & 105 & 50 & 35 \\
    \end{tabular}%
  \end{center}%
  \caption{Performance of Servo against Mozilla's Gecko rendering engine on the layout portion of some common sites.
  Times are in milliseconds, where lower numbers are better.}
  \label{servo-perf}
\end{table}

In the remainder of this section, we cover specific areas of Servo's design or implementation that make use of
Rust and the impacts and limitations of these features.

\subsection{Rust's syntax}
Rust has struct and enum types (similar to Standard ML's record types and datatypes~\cite{sml97-definition}) as
well as pattern matching.
These types and associated language features provide two large benefits to Servo over traditional browsers
written in \Cplusplus{}.
First, creating new abstractions and intermediate representations is syntactically easy, so there is very little
pressure to tack additional fields into classes simply to avoid creating a large number of new header and implementation
files.
More importantly, pattern matching with static dispatch is typically faster than a virtual function call on a class
hierarchy.
Virtual functions can both have an in-memory storage cost associated with the virtual function tables (sometimes many thousands of bytes\footnote{\url{https://chromium.googlesource.com/chromium/blink/+/c048c5c7c2578274d82faf96e9ebda4c55e428da}}) but more importantly
incur indirect function call costs.
All \Cplusplus{} browser implementations transform performance-critical code to either use the \lstinline[language=C]{final}
specifier wherever possible or specialize the code in some other way to avoid this cost.

Rust also attempted to stay close to familiar syntax, but did not require full fidelity or easy porting of
programs from languages such as \Cplusplus{}.
This approach has worked well for Rust because it has prevented some of the complexity that arose in
Cyclone~\cite{cyclone} with their attempts to build a safe language that required minimal porting effort
for even complicated C code.

\subsection{Compilation strategy}
Many statically typed implementations of polymorphic languages such as Standard ML of New Jersey~\cite{SMLNJ} and
\ocaml{}~\cite{ocaml-manual-3.0} have used a compilation strategy that optimizes representations of data types when
polymorphic code is monomorphically used, but defaults to a less efficient style otherwise, in order to share
code~\cite{ocaml-repr}.
This strategy reduces code size, but leads to unpredictable performance and code, as changes to a codebase that
either add a new instantiation of a polymorphic function at a given type or, in a modular compilation setting, that
expose a polymorphic function externally, can change the performance of code that is not local to the change being
made.

Monomorphization, as in MLton~\cite{mlton-compiler}, instead instantiates each polymorphic code block at each of the types
it is applied against, providing predictable output code to developers at the cost of some code duplication.
This strategy is used by virtually all \Cplusplus{} compilers to implement templates, so it is proven and well-known
within systems programming.
Rust also follows this approach, although it improves on the ergonomics of \Cplusplus{} templates
by embedding serialized generic function ASTs within ``compiled'' binaries.

Rust also chooses a fairly large default compilation unit size. A Rust \textit{crate} is subject to whole-program
compilation~\cite{weeks:whole-program-mlton}, and is optimized as a unit. A crate may comprise hundreds of modules,
which provide namespacing and abstraction. Module dependencies within a crate are allowed to be cyclic.

The large compilation unit size slows down compilation and especially diminishes the ability to build code in parallel.
However, it has enabled us to write Rust code that easily matches the sequential speed of its \Cplusplus{} analog, without requiring the Servo developers to become compiler experts.
Servo contains about 600 modules within 20 crates.

\subsection{Memory management}
As described in \secref{sec:rust}, Rust has an affine type system that ensures every value is used at
most once.
One result of this fact is that in the more than two years since Servo has been under development, we have
encountered zero use-after-free memory bugs in safe Rust code.
Given that these bugs make up such a large portion of the security vulnerabilities in modern browsers,
we believe that even the additional work required to get Rust code to pass the type checker initially is
justified.

One area for future improvement is related to allocations that are not owned by Rust itself.
Today, we simply wrap raw C pointers in \lstinline[language=Rust]{unsafe} blocks when we need to use a
custom memory allocator or interoperate with the SpiderMonkey JavaScript engine from Gecko.
We have implemented wrapper types and compiler plugins that restrict incorrect uses of these foreign values,
but they are still a source of bugs and one of our largest areas of unsafe code.

Additionally, Rust's ownership model assumes that there is a single owner for each piece of data.
However, many data structures do not follow that model, in order to provide multiple traversal
APIs without favoring the performance of one over the other.
For example, a doubly-linked list contains a back pointer to the previous element to aid
in traversals in the opposite direction.
Many optimized hashtable implementations also have both hash-based access to items
and a linked list of all of the keys or values.
In Servo, we have had to use unsafe code to implement data structures with this form,
though we are typically able to provide a safe interface to users.

\subsection{Language interoperability}
Rust has nearly complete interoperability with C code, both exposing code to and using code from
C programs.
This support has allowed us to smoothly integrate with many browser libraries, which has been
critical for bootstrapping a browser without rewriting all of the lower-level libraries immediately, such as
graphics rendering code, the JavaScript engine, font selection code, etc.
\tblref{servo-loc} shows the breakdown between current lines of Rust code (including generated code that
handles interfacing with the JavaScript engine) and C code.
This table also includes test code, though the majority of that code is in HTML and JavaScript.
\begin{table}
  \begin{center}
    \begin{tabular}{r || r}
      Language & Lines of Code \\
      \hline
      C or \Cplusplus{} & 1,678,427 \\
      Rust & 410,817 \\
      HTML or JavaScript & 217,827 \\
    \end{tabular}%
  \end{center}%
  \caption{Lines of code in Servo}
  \label{servo-loc}
\end{table}

There are two limitations in the language interoperability that pose challenges for Servo today.
First, Rust cannot currently expose varargs-style functions to C code.
Second, Rust cannot compile against \Cplusplus{} code.
In both cases, Servo uses C wrapper code to call into the code that Rust cannot directly
reach.
While this approach is not a large problem for varargs-style functions, it defeats many of the
places where the \Cplusplus{} code has been crafted to ensure the code is inlined into the caller,
resulting in degraded performance. We intend to fix this through cross-language inlining, taking
advantage of the fact that both \lstinline{rustc} and \lstinline{clang++} can produce output in the
LLVM intermediate representation~\cite{LLVM}, which is subject to link-time optimization. We have
demonstrated this capability at small scale, but have not yet deployed it within Servo.

\subsection{Libraries and abstractions}
Many high-level languages provide abstractions over I/O, threading, parallelism, and concurrency.
Rust provides functionality that addresses each of these concerns, but they are designed as thin
wrappers over the underlying services, in order to provide a predictable, fast implementation
that works across all platforms.
Therefore, much like other modern browsers, Servo contains many of its own specialized implementations of
library functions that are tuned for the specific cases of web browsers.
For example, we have special ``small'' vectors that allow instantiation with a default inline size,
as there are use cases where we create many thousands of vectors, nearly none of which have more than 4~elements.
In that case, removing the extra pointer indirection --- particularly if the values are of less than pointer size ---
can be a significant space savings.
We also have our own work-stealing library that has been tuned to work on the
DOM and flow trees during the process of styling and layout,
as described in \secref{sec:browsers}.
It is our hope that this code might be useful to other projects as well, though it is fairly browser-specific today.

Concurrency is available in Rust in the form of CML-style channels~\cite{reppy:cml}, but with a separation
between the reader and writer ends of the channel.
This separation allows Rust to enforce a multiple-writer, single-reader constraint, both simplifying and improving
the performance of the implementation over one that supports multiple readers.
We have structured the entire Servo browser engine as a series of threads that communicate over channels,
avoiding unsafe explicitly shared global memory for all but a single case (reading properties in the flow tree
from script, an operation whose performance is crucially tested in many browser benchmarks).
The major challenge we have encountered with this approach is the same one we have heard from other designers
of large concurrent systems --- reasoning about whether protocols make progress or threads eventually terminate
is manual and quite challenging, particularly in the presence of arbitrary thread failures.

\subsection{Macros}

Rust provides a hygienic macro system. Macros are defined using a declarative, pattern-based syntax~\cite{Kohlbecker:1987:MDS:41625.41632}. The macro system has proven invaluable; we have defined more than one hundred macros throughout Servo and its associated libraries.

\begin{figure}
\begin{lstlisting}
match self.state {
  states::Data => loop {
    match get_char!(self) {
      '&'  => go!(self: consume_char_ref),
      '<'  => go!(self: to TagOpen),
      '\0' => go!(self: error; emit '\0'),
      c    => go!(self: emit c),
\end{lstlisting}
  \caption{Incremental HTML tokenizer rules, written in a succinct form using macros. Macro invocations are of the form \lstinline{identifier!(...)}.}
  \label{fig:tokenizer-macros}
\end{figure}

For example, our HTML tokenizer rules, such as those shown in \figref{fig:tokenizer-macros}, are written in a macro-based domain specific language that closely matches the format of the HTML specification.\footnote{\url{https://html.spec.whatwg.org/multipage/syntax.html#tokenization}} Another macro handles incremental tokenization, so that the state machine can pause at any time to await further input. If no next character is available, the \lstinline{get_char!} macro will cause an early return from the function that contains the macro invocation. This careful use of non-local control flow, together with the overall expression-oriented style, makes Servo's HTML tokenizer unusually succinct and comprehensible.

The Rust compiler can also load compiler plugins written in Rust. These can perform syntactic transformations beyond the capabilities of the hygienic pattern-based macros. Compiler plugins use unstable internal APIs, so the maintenance burden is high compared to pattern-based macros. Nevertheless, Servo uses procedural macros for a number of purposes, including building perfect hash maps at compile time,\footnote{\url{https://github.com/sfackler/rust-phf}} interning string literals, and auto-generating GC trace hooks. Despite the exposure to internal compiler APIs, the deep integration with tooling makes procedural macros an attractive alternative to the traditional systems metaprogramming tools of preprocessors and code generators.

\subsection{Project-specific static analysis}

Compiler plugins can also provide ``lint'' checks\footnote{\url{http://doc.rust-lang.org/book/plugins.html#lint-plugins}} that use the same infrastructure as the compiler's built-in warnings. This allows project-specific safety or style checks to integrate deeply with the compiler. Lint plugins traverse a typechecked abstract syntax tree (AST), and they can be enabled/disabled within any lexical scope, the same way as built-in warnings.

Lint plugins provide some essential guarantees within Servo. Because our DOM objects are managed by the JavaScript garbage collector, we must add GC roots for any DOM object we wish to access from Rust code. Interaction with a third-party GC written in \Cplusplus{} is well outside the scope of Rust's built-in guarantees, so we bridge the gap with lint plugins. These capabilities enable a safer and more correct interface to the SpiderMonkey garbage collector. For example, we can enforce at compile time that, during the tracing phase of garbage collection, all Rust objects visible to the GC will report all contained pointers to other GC values, avoiding the threat of incorrectly collecting reachable values. Furthermore, we restrict the ways our wrappers around SpiderMonkey pointers can be manipulated, thus turning potential runtime memory leaks and ownership semantic API mismatches into static compiler errors instead.

As the ``lint'' / ``warning'' terminology suggests, these checks may not catch all possible mistakes. Ad-hoc extensions to a type system cannot easily guarantee soundness. Rather, lint plugins are a lightweight way to catch mistakes deemed particularly common or damaging in practice. As plugins are ordinary libraries, members of the Rust community can share lint checks that they have found useful.\footnote{\url{https://github.com/Manishearth/rust-clippy}}

Future plans include refining our safety checks for garbage collected values, such as flagging invalid ownership
transference, and introducing compile-time checks for constructs that are non-optimal in terms of performance or
memory usage.

%% file: open.tex
%!TEX root = paper.tex

\section{Open problems}
\label{sec:open}
While this work has discussed many challenges in browser design and our current progress,
there are many other interesting open problems.

\paragraph{Just-in-time code.} JavaScript engines dynamically produce native code that is
intended to execute more efficiently than an interpreted strategy.
Unfortunately, this area is a large source of security bugs.
These bugs come from two sources.
First, there are potential correctness issues.
Many of these optimizations are only valid when certain conditions of the calling
code and environment hold, and ensuring the specialized code is called only when those
conditions hold is non-trivial.
Second, dynamically producing and compiling native code and patching it into memory
while respecting all of the invariants required by the JavaScript runtime (e.g., the
garbage collector's read/write barriers or free vs. in-use registers) is also a challenge.

\paragraph{Integer overflow/underflow.} It is still an open problem
to provide optimized code that checks for overflow or underflow without
incurring significant performance penalties.
The current plan for Rust is to have debug-only checking of integer ranges
and for Servo to run debug builds against a test suite, but that may miss
scenarios that only occur in optimized builds or that are not represented
by the test suite.

\paragraph{Unsafe code correctness.} Today, when we write unsafe code in Rust
there is limited validation of memory lifetimes or type safety within that
code block.
However, many of our uses of unsafe code are well-behaved translations of
either pointer lifetimes or data representations that cannot be annotated
or inferred in Rust.
We are very interested in additional annotations that would help us prove
basic properties about our unsafe code, even if these annotations require a
theorem prover or ILP solver to check.

\paragraph{Incremental computation.} As mentioned in \secref{sec:browsers},
all modern browsers use some combination of dirty bit marking and incremental
recomputation heuristics to avoid reprocessing the full page when
a mutatation is performed.
Unfortunately, these heuristics are not only frequently the source of
performance differences between browsers, but they are also a source of
correctness bugs.
A library that provided a form of self adjusting computation suited to
incremental recomputation of only the visible part of the page, perhaps
based on the Adapton~\cite{adapton} approach, seems promising.

%%% Local Variables: 
%%% mode: latex
%%% TeX-master: "paper"
%%% End: 

%% file: rel.tex
%!TEX root = paper.tex
%
\section{Related browser research}
\label{sec:rel}
The ZOOMM browser was an effort at Qualcomm Research to build a parallel browser, also
focused on multicore mobile devices~\cite{ZOOMM}.
This browser includes many features that we have not yet implemented in Servo, particularly
around resource fetching.
They also wrote their own JavaScript engine, which we have not done.
Servo and ZOOMM share an extremely concurrent architecture --- both have script, layout,
rendering, and the user interface operating concurrently from one another in order to maximize
interactivity for the user.
Parallel layout is one major area that was not investigated in the ZOOMM browser, but is
a focus of Servo.
The other major difference is that Servo is implemented in Rust, whereas ZOOMM is written
in \Cplusplus, similarly to most modern browser engines.

Ras Bodik's group at the University of California Berkeley worked on a parallel browsing
project (funded in part by Mozilla Research) that focused on improving the parallelism
of layout~\cite{parallel-layout}.
Instead of our approach to parallel layout, which focuses on multiple parallel tree
traversals, they modeled a subset of CSS using attribute grammars.
They showed significant speedups with their system over a reimplementation of Safari's
algorithms, but we have not used this approach due to questions of whether it is possible
to use attribute grammars to both accurately model the web as it is implemented today
and to support new features as they are added.
Servo uses a very similar CSS selector matching algorithm to theirs.